# A security approach based on honeypots: Protecting Online Social network from malicious profiles

Fatna Elmendili, Nisrine Maqran, Younes El Bouzekri El Idrissi, Habiba Chaoui

*Computer Sciences Systems Engineering Laboratory, National School of Applied Sciences, Ibn Tofail University, Kenitra, Morocco*

A B S T R A C T

*In the recent years, the fast development and the exponential utilization of social networks have prompted an expansion of social Computing. In social networks users are interconnected by edges or links, where Facebook, twitter, LinkedIn are most popular social networks websites. Due to the growing popularity of these sites they serve as a target for cyber criminality and attacks. It is mostly based on how users are using these sites like Twitter and others. Attackers can easily access and gather personal and sensitive user's information. Users are less aware and least concerned about the security setting. And they easily become victim of identity breach. To detect malicious users or fake profiles different techniques have been proposed like our approach which is based on the use of social honeypots to discover malicious profiles in it. Inspired by security researchers who used honeypots to observe and analyze malicious activity in the networks, this method uses social honeypots to trap malicious users. The two key elements of the approach are: (1) The deployment of social honeypots for harvesting information of malicious profiles. (2) Analysis of the characteristics of these malicious profiles and those of deployed honeypots for creating classifiers that allow to filter the existing profiles and monitor the new profiles.*

## 1. Introduction

Social networks are now part of our daily life, we have certainly several accounts said "social", which are related to our daily lives (Facebook, Twitter), our professional life (Viadeo or LinkedIn), our sporting life or associative (we can cite jogg.in) and why not on sites of meeting (Adopteunmec, Meetic, Lovoo tinder or).

We can divide social networks on those which do not promote the anonymity (Facebook, Viadeo, LinkedIn) or the others which are promoting the anonymity (of general way all the sites of meeting, but also of services of Visio-conference as Skype), we have certainly received requests for connections from persons completely unknown. These applications are, in most of the cases, issued by malicious profiles.

That they emanate from the robots or that they are created to spoof the identity of a user, the malicious profiles are in constant increase on the Internet. On social networks, the malicious profiles can be generated by machines or be the result of identity theft and their motivations are various: Spy a PERSON, Increase the number of fans of a page (Facebook, Twitter…), spammer friends in impunity, fit all types of scams (very often of blackmail), harm the reputation of a person or a company, etc.

In particular, social spammers are increasingly targeting those systems as part of phishing attacks. To disseminate malware, and commercial spam messages, and to promote affiliate websites [1]. In only the past year, more than 80% of social networking users have "received unwanted friend requests, messages, or postings on their social or professional network account". Unlike traditional email based spam, social spam often contains contextual information that can increase the impact of the spam (e.g., by eliciting a user to click on a phishing link sent from a "friend") [1].

Successfully defending against these social spammers is important to improve the quality of experience for community members, to lessen the system load of dealing with unwanted and sometimes dangerous content, and to positively impact the overall value of the social system going forward. However, few information is known about these social spammers, their level of sophistication, or their strategies and tactics. Filling this need is challenging, especially in social networks consisting of 100s of millions of user profiles (like Facebook, Myspace, Twitter, YouTube, etc.). Traditional techniques for discovering evidence of spam users often rely on costly human-in the-loop inspection of training data for building spam classifiers; since spammers constantly adapt their strategies and tactics, the learned spam signatures can go stale quickly. An alternative spam discovery technique relies on community-contributed spam referrals (e.g., Users A, B, and C report that User X is a spam user); of course, these kinds of referral systems can be manipulated themselves to yield spam labels on legitimate users, thereby obscuring the labeling effectiveness, and neither spam discovery approach can effectively handle zero-day social spam attacks for which there is no existing signature or wide evidence [1]. With these challenges in mind, we propose and evaluate a novel Honeypot-based approach for uncovering social spammers in online social systems. Concretely, the proposed approach is designed to (1) the deployment of social honeypots for harvesting information of malicious profiles [8, 9]. (2) Analysis of the characteristics of these malicious profiles and those of deployed honeypots for creating classifiers that allow to filter the existing profiles and monitor the new profiles [2]. Drawing inspiration from security researchers who have used honeypots to observe and analyze malicious activity (e.g., for characterizing malicious hacker activity ,generating intrusion detection signatures,),In This extended paper we deploy and maintain social honeypots for trapping evidence of spam profile behavior, so that users who are detected by the honeypot have a high likelihood of being a spammer [1].This paper is an extension of work originally presented in conference 2016 4th IEEE International Colloquium on Information Science and Technology (CIST) in Tangier Morocco, we describe the processes of the proposed approach, starting with the deployment of social honeypots, the use of both feature based strategy and honeypot feature based strategy methods for collecting data, and finally we give the results and the test of this approach by using a dataset of profiles in machine learning based classifiers for identifying malicious profiles [2].These results are quite promising and suggest that our analysis techniques may be used to automatically identify the malicious profiles in social network.

*Corresponding Author: Elmendili Fatna, Computer Sciences Systems Engineering Laboratory, National School of Applied Sciences, Ibn Tofail University, Kenitra, Morocco, Email:f.elmendili@gmail.com.

*2.1. Overall Framework*

Malicious profiles are increasingly targeting Web-based social systems (like Facebook, Myspace, YouTube, etc.) as part of phishing attacks, to disseminate malware and commercial spam messages, and to promote affiliate websites. Successfully defending against these social spammers is important to improve the quality of experience for community members, to lessen the system load of dealing with unwanted and sometimes dangerous content, and to positively impact the overall value of the social system going forward. However, little is known about these social spammers, their level of sophistication, or their strategies and tactics [3].

In our ongoing research, we are developing approach for uncovering and investigating malicious user. Concretely, the Approach for detecting malicious profiles based Social Honeypot to (1) the deployment of social honeypots for harvesting information of malicious profiles. (2) Analysis of the characteristics of these malicious profiles and those of deployed honeypots for creating classifiers that allow to filter the existing profiles and monitor the new profiles [3]. Drawing inspiration from security researchers who have used honeypots to observe and analyze malicious activity. The Approach for detecting malicious profiles based Social Honeypot deploys and maintains social honeypots for trapping evidence of malicious profile behavior. In practice, we deploy a social honeypot consisting of a legitimate profile and an associated bot to detect social spam behavior. If the social honeypot detects suspicious user activity (e.g., the honeypot's profile receiving an unsolicited friend request) then the social honeypot's bot collects evidence of the spam candidate (e.g., by crawling the profile of the user sending the unsolicited friend request plus hyperlinks from the profile to pages on the Web-at-large). What entails suspicious user behavior can be optimized for the particular community and updated based on new observations of spammer activity [3].

While social honeypots alone are a potentially valuable tool for gathering evidence of social spam attacks and supporting a greater understanding of spam strategies, the goal of the Approach for detecting malicious profiles based Social support ongoing and active automatic detection of new Honeypot is to and emerging spammers. As the social honeypots collect spam evidence, we extract observable features from the collected candidate spam profiles (e.g., number of friends, text on the profile, age, etc.). Coupled with a set of known legitimate (non-spam) profiles which are more populous and easy to extract from social networking communities, these spam and legitimate profiles become part of the initial training set of a spam classifier [3].

As the social honeypots collect evidence on the Malicious Behaviors, the characteristics of profiles are extracted from the data of malicious profiles (for example: number of friends, Text, on the Profile, the age, etc.). Coupled to a set of legitimate profiles which are easy to extract the communities of social networks .This is called type of strategy by "Feature based strategy."[2].

A new method used in our approach to improve our classification and increase the ability to detect an attacker on the social networks that is "honeypot feature based strategy", this strategy uses the whole of char-Characteristics of honeypots that interact with users to refine our ranking [2].

The whole of the data collected are becoming an integral part for the training of a classifier of malicious profiles. By iterative refinement of selected characteristics using a set of algorithms for automatic classification which are implement on "Weka Machine Learning Toolkit" we can explore the more wide space of malicious profiles[14]. Fig. 1 present the approach for detecting malicious users.

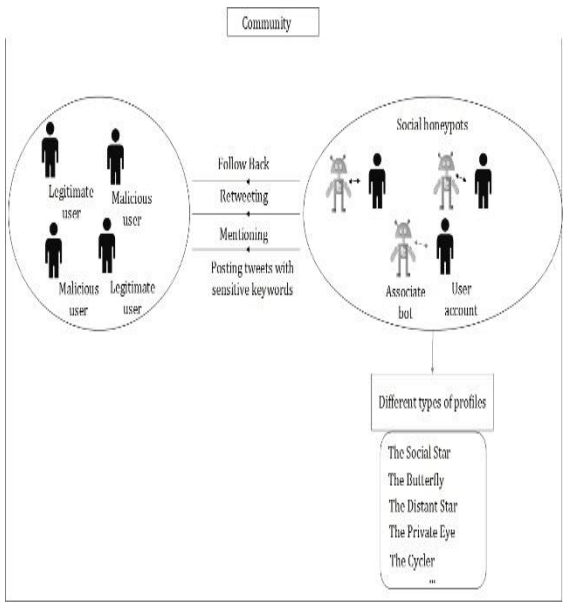

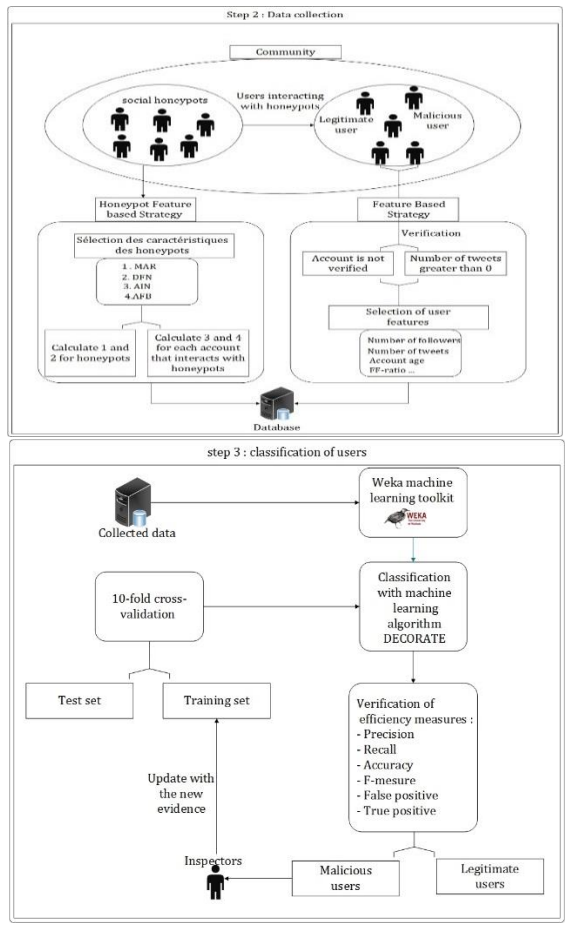

Fig.1 Overall Framework [2].

## 2. Malicious Profiles Detection Results:

In this section we present the results of our experiment. Creating of social honeypots: To develop our approach, we created 20 honeypots on twitter to trap malicious users and we analyze all the characteristics of users and deployed honeypots using weka that provides a platform algorithms of artificial intelligence and machine learning including Decorate algorithm that we used in our approach. We investigated various characteristics of the 90 friend requests that we collected with our social honeypots.

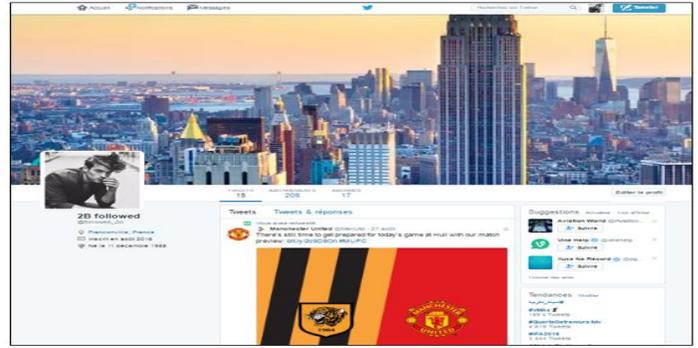

Fig. 2: Example of created honeypots

### 2.1. Development Tools:

- ***Twitter***

Twitter has become in the space of a few months a media phenomenon on the Internet. Everyone is put to talk in good or bad but without ever remain indifferent to them. Pushed by a few influential on the web, this small service extends more and faster its community.

Twitter has become in the space of a few months a media phenomenon on Internet. Everyone is put to talk in good or bad but without ever remain indifferent to them. Pushed by a few influential on the web, this small service extends more and faster its community.

Twitter is a tool managed by the enterprise Twitter Inc. It allows a user to send free short messages, known as tweets, on the Internet, by instant messaging or by SMS. Those messages are limited to 140 characters. The concept was launched in March 2006 by the company obvious based in San Francisco. The service rapidly became popular, up to bring together more than 500 million users in the world at the end of February 2012. At the 6 May 2016, twitter account 320 million active users per month with 500 million tweets sent by day and is available in more than 35 languages [4].

The first objective of Twitter and the reason of its deployment is to be able to provide a simple answer to the question: what am I doing? The use is very simple and free: We have 140 characters to disseminate our messages to whoever wants to receive if we specify our account in public or to our network only if it is in private. At the same time, we choose the members of Twitter which we want to follow the publications and these Members can we also add in return in their own network.

To publish our messages, several means are available: via the Twitter website, via our mobile phone by SMS, via Instant messaging type Google Talk or via software/third party Web sites based on the api free to twitter [5].

- ***Weka machine learning toolkit:***

Weka (for Waikato environment for knowledge Analysis) is a tool for data search open-source (GNU license) developed in Java. It was created at the University of Waikato New Zealand, by a group of researchers from the automatic learning, recognition of forms and the search of data. The software allows you to deal with different sources of data: files of various formats, including the format Attribute-Relation File Format (RTTW), developed for Weka; URLs; SQL databases. The analysis can be performed using most of the techniques of existing search.

The bibliographic reference attached to the Software is the book: data mining, practical machine learning tools and techniques with Java implementations, Ian H. Witten & Eibe Frank[6].

*2.2. Classification results:*

- ***Collection of data:***

After implementing our honeypots and interact with different types of users, we selected 90 profiles among 300 profile strapped by our honeypots and for each profile we selected traditional characteristics «traditional features" such as (Follower number, FF-ratio, Account age, Tweet Number, Mention ratio, ratio URL ...) and features based on honeypots "honeypot based features" such as (the number of honeypots with whom one interacts account, the daily average of new followers fora honeypot ...). The size of the database is 90 profiles.

TABLE 1: Dataset of users

Further, the data is converted to ARFF (Attribute Relation File Format) format to process in WEKA. An ARFF file is an ASCII text file that describes a list of instances sharing a set of attributes. User Classification with Weka: After preparing our data file, we will use the classification algorithms implemented in Weka for test methods, solve problems, focusing us on the use and the results provided by these implementations without having to rewrite every time algorithms.

The framework within which we will work, and algorithms that we will study and use are based on the following functional diagram:
-There is a set of examples, each example being defined by Description: This is a set of values defining this example. The class that he was associate (with the help of a human expert) [7].

-This set of examples is provided to a program that will generate a classifier. A classifier is a program that, when provided him an example, try to guess its class. In other words, the program tries to guess the class of an instance from its description. After processing the ARFF file in WEKA the list of all attributes, statistics and other parameters can be utilized as shown in Fig. 3.

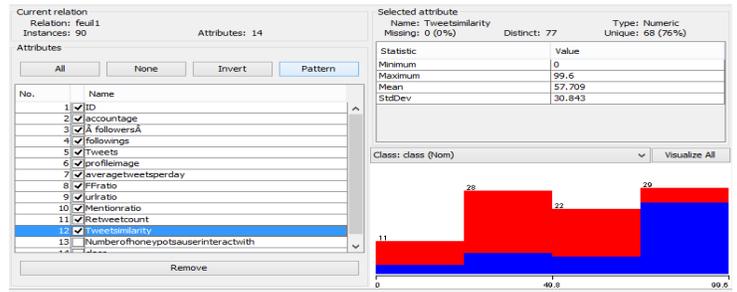

Fig. 3: Processed ARFF file in WEKA

In the above shown file, there are 90 profiles data is processed with different attributes like followers, FF ratio, Retweet-count, Mention-ratio, etc. [10,11,12].

The processed data in Weka can be analyzed using different data mining techniques like, Classification, Clustering, Association rule mining, Visualization, etc. The Fig. 4 shows the few processed attributes which are visualized into a 2 dimensional graphical Representation.

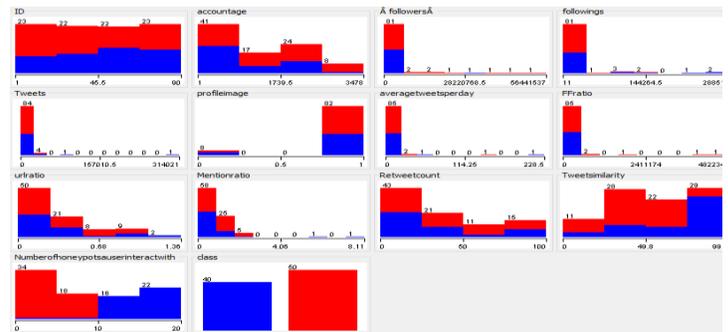

Fig. 4: Graphical visualization of processed attributes

Now that we have loaded our dataset, we used DECORATE machine learning algorithm to model the problem and make predictions and we choose Cross-validation, which lets WEKA build a model based on subsets of the supplied data and then average them out to create a final model[15].

In previous work, generally the métas Classifiers (Decorate, Lo-gitBoost, etc.) Product of Best performance that the classifiers to trees (BFTree and FT) and classifiers based on functions (SimpleLogistic and libsvm) [7]. For our approach we chose decorate as classifier. For different reasons:
-The speed and the execution time;
-Best Performance Compared to Other;
-Of Results approximately correct.

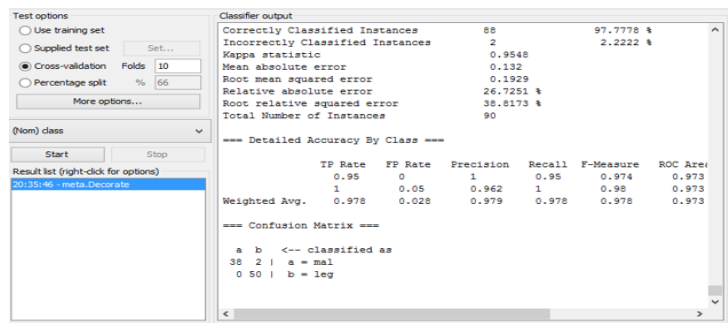

Fig. 5: Classifier output

Fig.5 shows estimates of the trees predictive performance, generated by WEKAs evaluation module [16]. It outputs the list of statistics summarizing how accurately the classifier was able to predict the true class of the instances under the chosen test module. The set of measurements is derived from the training data. In this case 97.7778% of 90 training instances have been classified correctly. This indicates that the results obtained from the training data are optimistic compared with what might be obtained from the independent test set from the same source. In addition to classification error, the evaluation output measurements derived from the class probabilities assigned by the tree. More specifically, it outputs mean output error (0.132) of the probability estimates, the root mean squared error (0.1929) is the square root of the quadratic loss. Theme an absolute error calculated in a similar way by using the absolute instead of squared difference. The reason that the errors are not 1 or 0 is because not all training instances are classified correctly. Kappa statistic is a chance-corrected measure of agreement between the classifications and the true classes. It's calculated by taking the agreement expected by chance away from the observed agreement and dividing by the maximum possible agreement. The Kappa coefficient is calculated as follows:

$$K = \frac{P_o - P_e}{1 - P_e} \quad (1)$$

With $P_0$ : is the proportion of the sample on which the two judges are of the agreement.
(i.e. the main diagonal of the matrix of confusion).
And $p_e = \frac{\sum_i p_i \, p.i}{n^2} \quad (2)$

Where:

$p_i$ is the sum of the elements of the line i;
p.i is the sum of the elements of the COLUMN I;
n: in a sample size.

The Kappa coefficient takes values between -1 and 1:
- It is maximal when both judgments are the same:

All examples are on the diagonal, and P0 = 1.
- It is 0 when both judgments are independent (P0=Pe).

- It is -1 when the judges disagree. Detailed Accuracy by Class demonstrates a more detailed per class break down of the classifiers prediction accuracy.

- The True Positive (TP) rate is the proportion of examples which were classified as class x, among all examples which truly have class x, i.e. how much part of the class was captured. It is equivalent to Recall. In the confusion matrix, this is the diagonal element divided by the sum over the relevant row, **i.e.** 38/ (38+2) =0.95 for class malicious and 50/ (50+1) =1 for class legitimate in our example.

- The Precision is the proportion of the examples which truly have class x among all those which were classified as class x. In the matrix, this is the diagonal element divided by the sum over the relevant column, **i.e.** 38/ (38+0) =1 for class malicious and 50/ (2+50)

=0.962 for class legitimate. From the Confusion matrix in Fig. 6 we can see that two instances of a class "legitimate" have been assigned to a class" malicious", and zero of class "malicious" are assigned to class "legitimate".

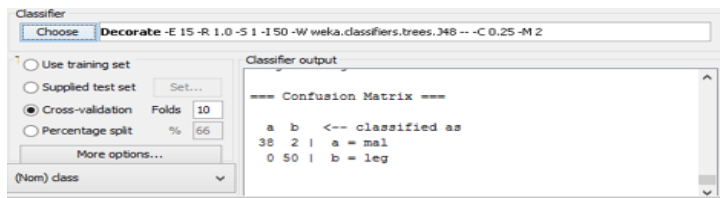

Fig. 6: Confusion matrix

Fig. 7 present the threshold curve for the prediction. This shows a 97.28% predictive accuracy on the malicious class.

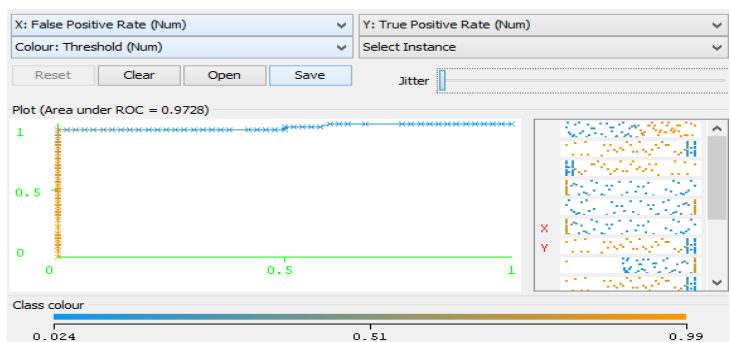

Fig. 7: Threshold curve

In order to find the optimal value of the threshold, we perform the cost/benefit analysis. Consider attentively the window for the Cost/Benefit Analysis. It consists of several panels. The left part of the window contains the Plot: Three should Curve frame with the Threshold Curve (called also the Lift curve).However, the axis X in the Threshold curve corresponds to the part of selected instances (the Sample Size). In other words, the Threshold curve depicts the dependence of the part of active compounds retrieved in the course of virtual screen in gup on the part of compounds selected from the whole data set used for screening. The confusion matrix for the current value of the threshold is shown in the Confusion Matrix frame at the left bottom corner of the window. We observed that the confusion matrix for the current value of the threshold sharply differs from the previously obtained one. In particular, the classification accuracy 74.4444% is considerably less than the previous value 97.7787%, the number of false positives has greatly increased from 0 to 21.

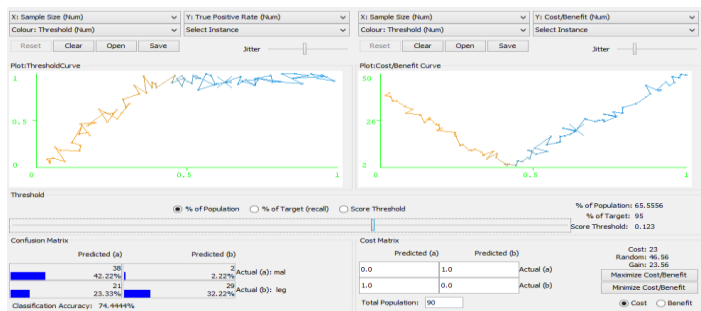

Fig. 8: cost/benefits analysis

In Fig.9, we generated a plot illustrating the prediction margin; the margin is defined as the difference between the probabilities predicted for the actual class and the highest Probability predicted for the other classes.

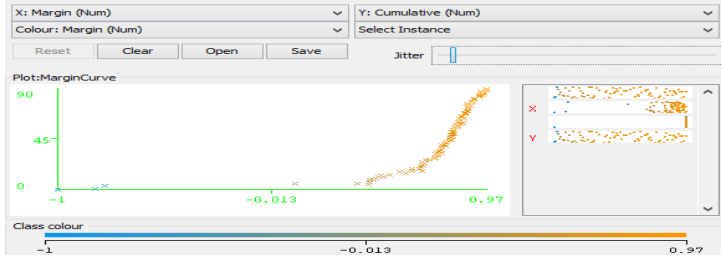

Fig. 9: Margin curve

In Fig.10 we present the performance of our malicious users detector trained with traditional features, honeypot based features and two sets of features together, respectively.

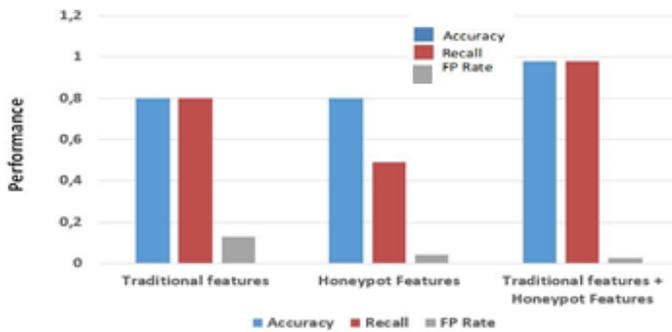

Fig. 10: Malicious user's detector performance with different feature set

We can find that after we combine traditional feature set with honeypot based feature set, we can achieve an accuracy of 0.979, a recall of 0.978 and a false positive rate of 0.028. The accuracy and recall are much better than simply using the other two feature sets independently. Though the FP rate is higher than simply using honeypot based feature set, we can modify the threshold to make a trade-off between FP rates and recall [2].

From the results obtained we can point out that with the hybrid method used (traditional feature set with honeypot based feature), we can detect more wide space of malicious users on social networks and why not apply the same approach to other communities. This hybrid approach gives results relevant to other methods.

3. Conclusion

In this paper, we have presented the results of a novel social honeypot-based approach to detect malicious profiles in social networking communities already published [2]. Our overall research goal is to investigate techniques and develop effective tools for automatically detecting and filtering malicious users who target social systems. Specifically, our approach deploys social honeypot profiles in order to attract malicious accounts. By focusing on twitter community, we use a set of user's characteristics and honeypots deployed characteristics to create a malicious profiles classifier based on machine learning algorithm Decorate for identifying malicious accounts with high precision and allow rate of false positives. In our ongoing work, we are using our analysis results to automatically identify malicious users in twitter social network. We have tried to apply our proposed approach using Weka to classify a set of users caught by our honeypots. From the results obtained we can point out that with the method of classification used, we can detect the more wide space of malicious users on social networks and why not apply the same approach to other communities. We hope this will give us a more macroscopic picture of the privacy awareness of general OSN (Online social network) users [13], and we want to use this to raise the awareness of privacy not only from the user sides but also for the OSN designers. Together with our research on detection malicious profiles in social network, we hope being able to contribute on making OSNs a safer place for the ordinary users [17, 18].